\documentclass[preprint,journal]{vgtc}                % final (journal style)
\ifpdf%                                % if we use pdflatex
  \pdfoutput=1\relax                   % create PDFs from pdfLaTeX
  \usepackage{graphicx}                % allow us to embed graphics files
  \DeclareGraphicsExtensions{.pdf,.png,.jpg,.jpeg} % for pdflatex we expect .pdf, .png, or .jpg files
\else%                                 % else we use pure latex
  \usepackage{graphicx}                % allow us to embed graphics files
  \DeclareGraphicsExtensions{.eps}     % for pure latex we expect eps files
\fi%
\graphicspath{{figure/}{pictures/}{images/}{./}} % where to search for the images
\usepackage{microtype}                 
\PassOptionsToPackage{warn}{textcomp}  
\usepackage{textcomp}               
\usepackage{mathptmx}                 
\usepackage{times}                 
      
\usepackage{cite}                 
\usepackage{tabu}                    
\usepackage{booktabs}             
\usepackage{xcolor}

\newcommand{\zhaosong}[1]{\textcolor{black}{#1}}

\graphicspath{{figures/}{pictures/}{images/}{./}} % where to search for the images
\usepackage{tabu}                      % only used for the table example
\usepackage{booktabs}                  % only used for the table example
\usepackage{multirow}
\usepackage{mathptmx}
\usepackage{times}
\usepackage{amsmath}
\usepackage{times}
\usepackage{epsfig}
\usepackage{psfrag}
\usepackage{amsmath}
\usepackage{amssymb}
\usepackage{url}
\usepackage[linesnumbered,boxed]{algorithm2e}
\usepackage[none]{hyphenat}     % No hyphens
\usepackage{color}
\newcounter{lenumerate}

\renewenvironment{itemize}{\begin{list} {\labelitemi}
{\setlength{\parsep}{0.1ex} \setlength{\topsep}{0.6ex}
\setlength{\partopsep}{0.1ex } \setlength{\itemsep}{0.2ex}
\setlength{\leftmargin}{2ex} }} {\end{list}}

\onlineid{1064}
\vgtccategory{Research}
\vgtcpapertype{Algorithm/Technique}

\ieeedoi{10.1109/TVCG.2019.2934671}

\title{A Natural-language-based Visual Query Approach\\ of Uncertain Human Trajectories}

\author{Zhaosong~Huang,
        Ye~Zhao,
        Wei~Chen,
        Shengjie~Gao,
        Kejie~Yu,
        Weixia~Xu,
        Mingjie~Tang,
        Minfeng~Zhu,
        and~Mingliang Xu}
\authorfooter{
\item Z. Huang, W. Chen, S. Gao, K. Yu, W. Xu, and M. Zhu are with The State Key Lab of CAD \& CG, Zhejiang University, China. E-mail: \{zhaosong\_huang, gaoshengjie, minfeng\_zhu\}@zju.edu.cn, chenwei@cad.zju.edu.cn, \{ykjage, xuweixia96\}@gmail.com. (W. Chen and M. Xu are corresponding authors)
\item Y. Zhao is with the Department of Computer Science, Kent State University, Kent, OH 44242, USA. E-mail: zhao@cs.kent.edu.
\item M. Tang is with the Ant Financial, USA. E-mail: tangrock@gmail.com.
\item M. Xu is with the School of Information Engineering, Zhengzhou University, Zhengzhou, 450000, China. E-mail: iexumingliang@zzu.edu.cn.
}
\shortauthortitle{Huang \MakeLowercase{\textit{et al.}}: A Natural-language-based Visual Query Approach}

\abstract{Visual querying is essential for interactively exploring massive trajectory data. However, the data uncertainty imposes profound challenges to fulfill advanced analytics requirements. On the one hand, many underlying data does not contain accurate geographic coordinates, e.g., positions of a mobile phone only refer to the regions (i.e., mobile cell stations) in which it resides, instead of accurate GPS coordinates. 
On the other hand, domain experts and general users prefer a natural way, such as using a natural language sentence, to access and analyze massive movement data.
In this paper, we propose a visual analytics approach that can extract spatial-temporal constraints from a textual sentence and support an effective query method over uncertain mobile trajectory data. 
It is built up on encoding massive, spatially uncertain trajectories by the semantic information of the POIs and regions covered by them, and then storing the trajectory documents in text database with an effective indexing scheme. The visual interface facilitates query condition specification, situation-aware visualization, and semantic exploration of large trajectory data. Usage scenarios on real-world human mobility datasets demonstrate the effectiveness of our approach.
}

\teaser{
 \centering
 \setlength{\abovecaptionskip}{0pt}
 \includegraphics[width=0.95\linewidth]{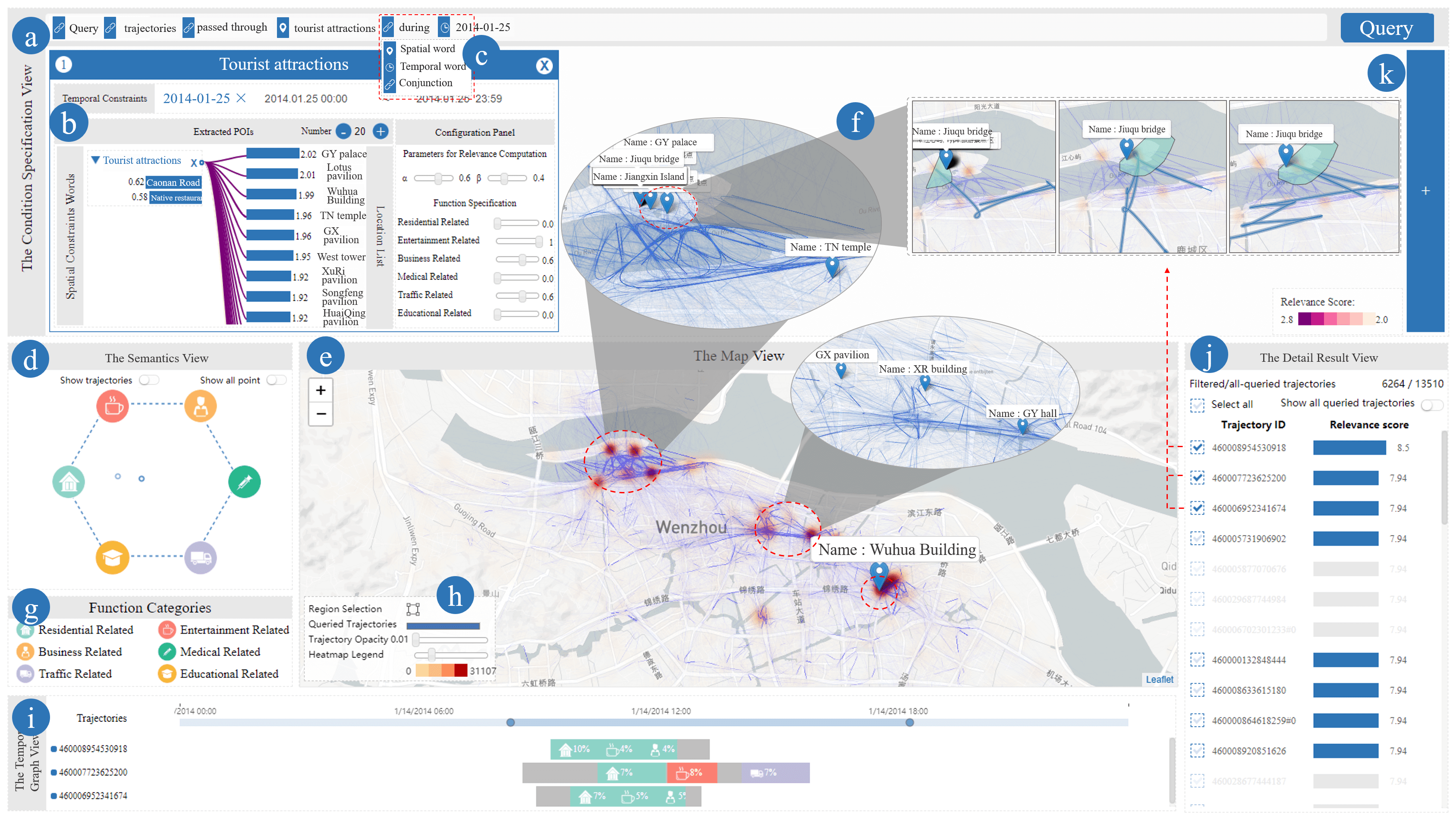}
 \caption{(a) The query condition specification view. (b) The relevance tree with the spatial keyword `tourist attractions'. (c) The drop-down menu for changing the type of the input keyword. (d) The semantics view shows that the major region functional topic is `residential related'. (e) The map view shows that the queried trajectories are mainly distributed in the northwest (named `Jiangxin island'), the middle, and the east (named `Wuhua building') of the city. (f) Most trajectories land the island from its east through ferry. (g) The region functional topics. (h) The rendering parameter widget. (i) The temporal graph view. (j) The detail result view of the queried trajectories. (k) Detail study of urban areas.
 }
 \vspace{-3mm}
 \label{fig:teaser}
}

\begin{document}

\firstsection{Introduction}\label{sec:introduction}

\maketitle
%Please use the following command to revise the paper: \textbackslash zhaosong\{test\}: \zhaosong{test}, \textbackslash mingjie\{test\}:\mingjie{test}, \textbackslash ye\{test\}:\ye{test}., \textbackslash wei\{test\}: \wei{test}.

A common task in analyzing massive trajectory data is to query the trajectories with given spatiotemporal conditions.
This has been proven useful to improve people's life quality~\cite{markovic2018applications}, urban planning~\cite{Yuan:2012:Discovering}, and real-time monitoring~\cite{zeng2013visualizing, wang2017adaptively, lv2018crowd}. 
A crucial problem in this process is to help data analysts express their query requirements intuitively and effectively, where interactive visual interfaces~\cite{chen2018vaud, zhou2019visual} can play an important role. However, specifying complex conditions in spatial and temporal dimensions is neither natural nor intuitive for domain users and practitioners, which can easily hinder their intention of utilizing the systems.

\begin{figure}[tb]
 \centering
 \includegraphics[width=0.9\columnwidth]{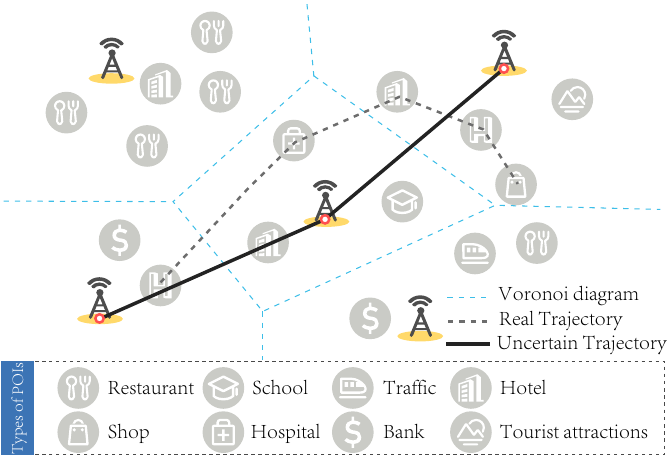}
 \caption{Illustration of the spatial uncertainty of a  mobile phone trajectory.}
 \label{fig:region}
\end{figure}

Using natural language input is undoubtedly a preferred way to express query conditions, where analyzers can naturally use location names (e.g., Golden Gate Bridge), functional categories (e.g., education areas, residential areas), and time descriptions (e.g., morning) to filter massive trajectories. The ``textualization-and-query'' scheme has been applied by querying geospatial data after externalizing locations with contextual geo-information (a.k.a., textualization). The annotated trajectory data provides knowledge-based context~\cite{Spaccapietra2008A} which dramatically enriches the raw data~\cite{Alvares2007A, Yuan:2012:Discovering, Richter2012Semantic, Hu2016Fuzzy}. This scheme is used in several visual analytics systems for studying taxi trajectories~\cite{Zhao:2014:TaxiTopics, Aldohuki2016SemanticTraj} and human mobility patterns~\cite{Zeng:2017:Visualizing}. \zhaosong{Most existing methods assume that the trajectories contain accurate geo-locations. However, many real-world trajectory datasets incorporate \textit{spatial uncertainty} due to (1) the inaccuracy and measurement error of sensors~\cite{fisher1999models}. GPS samples can divert from real positions while some samples along a trajectory may be missing; and (2) the location privacy protection~\cite{duckham2006location}. Data providers may hide an accurate point of longitude and latitude with a spatial unit (e.g., street, region) it resides.} Figure~\ref{fig:region} illustrates human trajectories sampled by mobile phones, where each data point only tells us the base station that the phone is connected to. The dashed line denotes a real human trace which is not achievable, while the solid line links the stations to form a trajectory that approximates the real trace. 
\zhaosong{The spatially uncertain trajectories have been proved essential in many application scenarios~\cite{andrienko2017visual} such as identification of commute patterns~\cite{wang2017adaptively, zhou2019visual}, analyzing people's activities~\cite{Chen2016Interactive, kruger2014semantic}, and spatial planning~\cite{liu2016smartadp}.
Similar to certain trajectories, the analytical goal is to quickly retrieve the trajectories based on spatiotemporal conditions and then discover patterns from the query results. 
However, spatially uncertain trajectories make it hard to access the data by specifying accurate locations. Therefore, querying them by a descriptive sentence such as ``trajectories passing a school'' becomes necessary and useful for analyzers. 
Existing textualization and query methods are not directly applicable since the sampling locations are not mapped to a fixed street name or POI  (Point of Interest, which is a specific point location, e.g., hotel.), which motivates us to develop new approach that integrates natural language based query with uncertain trajectory data.}

%In such cases, 

In this paper, we develop a new visual analytics system to fill the gap, which enables users to query spatially uncertain human trajectories with natural language sentences. For example, an urban planner can specify: `Query human trajectories passed through tourist attractions in the morning' for discovering mobility patterns of tourists and improving public services. \zhaosong{The system further provides a set of visual representations to help users easily study the returned trajectories. Our approach integrates high-level descriptive languages of locations and time. A trajectory query engine is developed to address two major challenges: the spatial uncertainty and inaccurate descriptive language. In this paper, we use the mobile phone trajectory data to describe our work, which can also be extended to handle other types of uncertain human trajectories (Sec. \ref{sec:discussion}).}

First, we understand and extract the conditions from a natural language query sentence. The spatiotemporal constraints are not given accurately. \zhaosong{A location name often includes entry error~\cite{fisher1999models} and have vague meanings.} For example, `school, college, campus, and university' may all be linked to one given word `school'. We address the challenge by hiring a natural language lexical analyzer to split and encode spatial and temporal expressions, \zhaosong{and then developing a query condition specification tool to visually identify and refine relevant words, where a word embedding model is trained to discover word relevance.}

Second, we query trajectories by the extracted conditions. \zhaosong{First, the city space is subdivided by small possible spatial regions (PSR) to accommodate inaccurate sample points.} Second, the trajectories are converted to documents containing POI information in the regions and then stored in a database with a special indexing scheme. Third, top-K POIs matching the input conditions are found by a probabilistic retrieval algorithm (Okapi BM25~\cite{robertson2009probabilistic}), which is further enhanced by utilizing the functional topics of regional POIs discovered by Latent Dirichlet Allocation (LDA~\cite{Blei2003Latent}) topic modeling. Eventually, the top-K POIs are used to extract relevant trajectories in the database. 

Furthermore, once a group of trajectories is acquired, we further rank them by a relevance score computed for each trajectory. We present an ordered list of the results which helps users efficiently investigate the trajectories that meet their criteria. 

A visual analytics system is built up with (1) a query condition specification view to specify a query and visually adjust query conditions.
(2) A map view to visualize trajectories within geo-context; and (3) a set of visualizations including a temporal graph view, a semantics view, and a detail result view for examining the query results.

Our contributions are summarized as follows:
\begin{itemize}
    \item An efficient \zhaosong{query engine of spatially uncertain human trajectories which is built upon natural language queries and textualization of the trajectories.}
    \item A multi-faceted visual interface designed for interactively specifying semantic query conditions and investigating retrieved uncertain trajectories.
    \item \zhaosong{Two real-world usage scenarios based on massive smartphone based human trajectories that demonstrate the usefulness of our approach.}
\end{itemize}

\section{Related Work}\label{sec:relatedwork}
\zhaosong{The goal of visual query is to retrieve and study trajectories by addressing the challenges of the large data volume and the diversity of query tasks~\cite{hendawi2012predictive,marsit2005query}.} Here we introduce relevant literature on query conditions, trajectory data queries, and semantic-based visual exploration of spatial trajectory data.

%2.1 包含两块完全不同的内容，建议分开或者标题里去掉trajectory data management。 因为这个不是论文的焦点
\subsection{Query Condition Specification}
Retrieving trajectories via programming languages~\cite{zheng2014urban} is widely used in data analysis systems designed for programmers instead of casual users.
Recently, visual query technology has been proven to be useful in helping users express their query requirements, such as defining spatial constraints on map~\cite{zhou2019visual, wang2014visual}, specifying Origin-Destination (OD) query~\cite{ferreira2013visual}, and performing road navigation~\cite{lu2017visual}.
Visual query systems also allow users to modify their query input by well-designed interactions.
For example, VAUD~\cite{chen2018vaud} helps users interactively construct and optimize data query conditions.
VESPa~\cite{haag2016vespa} employs a sketch-based interface for users to express, check, and refine hypotheses.

However, for complex data query tasks, users often need to conduct cumbersome interactions to refine the conditions which are neither natural nor intuitive.
Using natural language in visual queries has been shown as a natural way for user interaction.
Alvares et al.~\cite{Alvares2007A} propose a generic model to enrich trajectories with semantic geographical information and support semantic queries and analysis. SemanticTraj~\cite{Aldohuki2016SemanticTraj} fuses GPS sampling points of trajectories with their street and POI semantics and enables querying via textual input of street/POI names.
\zhaosong{However, the work does not support users to specify a fuzzy yet more natural query sentence (e.g., `Query trajectories of students').
Our approach overcomes this problem by extracting query constraints from natural language sentences and supporting visual condition inspection and optimization.}

\subsection{Efficient Trajectory Queries}
Spatial indexing is widely used for trajectory data management and supporting effective trajectory retrieval.
Feng et al.~\cite{feng2016survey} summarize seven kinds of queries, which mainly focuses on two categories~\cite{zheng2015trajectory}: Range queries and K-Nearest-Neighborhoods (KNN) queries. 

\textbf{Range queries} refer to filtering the trajectories falling into a spatial region. The tree-based index~\cite{hendawi2012predictive} is one of the most popular indexing methods to support it. 
%Quad-tree~\cite{finkel1974quad} recursively partition the spatial space into four regions and store spatial-closed data on the same disk partitions to improve the access performance.
Mokbel and Chen et al.~\cite{mokbel2003spatio, chen2008benchmark} review and evaluate the performance of existing spatial-temporal access methods.
The space-time cube (STC)~\cite{kraak2003space} enables fast indexing and access by dividing the space-time space into uniformly sampled grids.
Extended variants of the STC (e.g., Nanocubes~\cite{lins2013nanocubes}) are widely employed for supporting multivariate spatial-temporal data exploration~\cite{li2005new,muthumanickam2019identification}, aggregation~\cite{andrienko2008spatio,Buchmuller2018MotionRugsVC}, and interactive analysis~\cite{li2018semantics, andrienko2013visual,andrienko2019analysis}.
\textbf{KNN-based queries} refer to retrieving top K trajectories which have the minimum distance to given locations. R-tree~\cite{Guttman1984R,Deng2011Trajectory}, KD-Tree~\cite{Guttman1984R,Deng2011Trajectory} and their extensions (e.g., ~\cite{Sathya2014A,Pfoser2000Novel,Nascimento1998Towards,Zhou2005Close}) are widely used to improve the performance of KNN-based queries. SETI~\cite{chakka2003indexing} indexes spatial and temporal dimensions separately by a two-level index structure. TrajStore~\cite{Cudremauroux2010TrajStore} divides trajectories into subtrajectories and stores them geographically and temporally near each other in the same disk block.
While necessary, the above indexing schemas focus on accessing data in spatial-temporal dimensions.
\zhaosong{In this paper, our query engine is built upon an efficient documentation-indexing scheme for the fast query of uncertain trajectories through a simple text-based input.}

\subsection{Semantic Visual Exploration of Trajectories}
There are a large number of studies on visual analysis of trajectory data. We refer the readers to recent literature surveys for more details ~\cite{andrienko2013visual,chen2015survey}. Existing works mainly focus on displaying spatial-temporal information of trajectories on a map and visual charts. However, human mobility data~\cite{smoreda2013spatiotemporal} are massive and have complex semantics. Understanding the semantics which contain enriched knowledge (e.g., behaviors of moving objects.) has been an important research task. \zhaosong{A survey~\cite{Parent:2013:Semantic} summarizes the semantic enrichment, knowledge extraction, and mobility analysis approaches for the trajectory data.}
Many approaches use data fusing techniques to supplement trajectories from multiple domain data~\cite{shen2018streetvizor}. The POI data is often used to discover different and temporally changing functions of urban regions~\cite{Yuan:2012:Discovering, krueger2014visual} which can further enrich the trajectory data. 
Similarly, Zeng et al.~\cite{Zeng:2017:Visualizing} characterize population movement by the subway transportation data and related POIs for individual behavior investigation.
\zhaosong{Moreover, by reformulating trajectories into an appropriate semantic form ~\cite{Zhao:2014:TaxiTopics}, the LDA topic models are utilized to extract implicit themes and visualize them.}

\zhaosong{To depict the spatiotemporal attributes of the trajectories, the geographical context information is incorporated into the trajectory visualization.} For instance, a node-link graph ~\cite{Huang2015TrajGraph} is constructed and employed to study relations among different regions. The trajectory flow from OD movement data is encoded with a spatiotemporal abstraction to reveal patterns and trends of mass mobility~\cite{Andrienko2016Revealing}. \zhaosong{Andrienko et al.~\cite{andrienko2017state} use a state transition graph to display the mobility behaviors where each state represents a semantic category of location (e.g., shop).} Other representative works are dedicated to the exploration of stop points~\cite{Chen2016Interactive,Yang2017Many,Zeng2016Visualizing, xu2017efficient}, trajectory lines~\cite{Liu2011Visual, xu2019crowd}, and regions~\cite{Li2013Attraction,zeng2013visualizing}.
Most of these works focus on trajectories with accurate geo-location. \zhaosong{Our approach instead seeks to represent and visualize semantics from spatially uncertain trajectory data.
The intention is that we should allow users to intuitively use the geographical semantics and to achieve an iterative trajectory data exploration process.}

%Iovan et al.~\cite{iovan2013moving} quantify the mobile trajectory data at the local and global level, then they propose a sampling method to reduce the redundant information.  

%Krueger et al~\cite{krueger2014visual} use Point of interests to enrich the trajectory data and analyze the trajectory based on a timeline.
%G{\"u}nther et al.~\cite{sagl2012visual} propose a visual analytic method to explore a sizeable mobile movement dataset and reveal the hidden movement patterns of human mobility. 

%The enriched semantic trajectory can be further used for enabling analytical operations such as querying or measuring the similarities among trajectories~\cite{Hu2016Fuzzy}. 

%There are many methods to manage the uncertain information and are surveyed by MacEachern et al.~\cite{maceachren2005visualizing}.
%\subsection{Mining Semantics from Trajectories}

\section{\zhaosong{National Language Based Query Engine}}
To support querying uncertain trajectories with high-level descriptive language, the engine is built up on several modules including: (1) an uncertain trajectory documentation and indexing module to represent and manage uncertain trajectories; (2) a constraint extraction module to process the input sentence and form specific spatial-temporal conditions; and (3) a relevance quantification method that computes the relevance scores of uncertain trajectories to the given conditions. 

\subsection{Uncertain Trajectory Documentation and Indexing} \label{sec:documentindexing}
\zhaosong{An uncertain trajectory has a series of inaccuracy locations instead of accurate longitudes and latitudes. Each such location typically can be considered residing in a \textit{possible spatial region} (PSR). For example, an inaccurate GPS point may locate in a circular region with a radius of dozens of meters. A human mobile sample point may reside in the close neighborhood of a base station (Figure~\ref{fig:region}). We then design and implement an uncertain trajectory documentation and indexing mechanism based on the PSR.}

\zhaosong{Using human mobile trajectory data as an example, given the distribution of base stations, the urban space can be partitioned into PSR regions.} As shown in Figure~\ref{fig:region}, a Voronoi partitioning~\cite{aurenhammer1991voronoi} algorithm yields a set of disjoint Voronoi cells of a city, which can be considered as PSRs of the inaccurate trajectory points inside the city. Each city POI (Point of Interest) uniquely belongs to a certain PSR region, and each region contains multiple POIs. 

\begin{figure}[tb]
 \centering
 \includegraphics[width=\columnwidth]{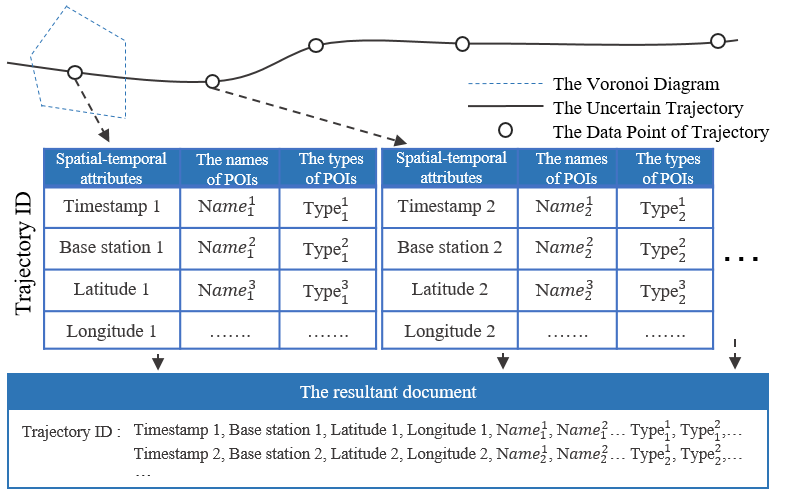}
 \caption{Trajectory documentation scheme of human mobile trajectories. Each uncertain data point is encoded with its timestamp, PSR information (base station ID and coordinates), and semantic context information. These information forms a trajectory document.}
 \label{fig:trajectorydocument}
\end{figure}

\noindent \textbf{Documentation:} An uncertain human trajectory is enriched with its contextual information and transformed into a text formulation. As shown in Figure~\ref{fig:trajectorydocument}, each trajectory data point contains a timestamp, its geographic PSR information (including the base station ID,  latitude and longitude), and its semantic information which are the names, types, and descriptions of the POIs inside its PSR. Then this information of all sample points on a trajectory forms a trajectory document. Following this process, all trajectories are converted into trajectory documents and stored into a text database.

\zhaosong{Please note that for different types of uncertain trajectory data, the PSRs can be computed in other ways, but the trajectory documents can be generated in the same way using the POIs inside the PSRs.}

\vspace{3pt}
\noindent \textbf{Indexing:} We propose a \textbf{temporal-textual-trajectory index} to support efficient semantic query. First, we divide the collection of trajectory documents into multiple data partitions based on a time interval, that is, trajectories staying in the same time window are fetched and stored into the same data partition. Then, trajectory documents in the same partition are stored into adjacent disk blocks to save disk I/O. Next, we build an inverted index for trajectory documents in the same data partition. This inverted index stores the IDs of trajectories whose documents contain the keywords of POIs names and types. Finally, we add the timestamps used for data partitions to generate a temporal-textual-trajectory index.

\noindent \textbf{Retrieval:} Given spatial and temporal query conditions, such as querying trajectories passing `Central Park' during `1:00 $pm$ to 2:00 $pm$'. We first fetch the data partitions that meet the temporal constraint. Because data partitions are sorted by time, a binary search is used to speed up this searching whose runtime overhead is $O(\log p)$, where $p$ is the total number of data partitions.
Then, for each qualified data partition, the inverted index identifies those trajectory documents that contain the keyword `Central Park' which involves $O(1)$ runtime cost for each partition. Our experiments show that this indexing scheme can accelerate trajectory retrieval with less latency for real-time exploration and visualization.

\subsection{Natural Language Processing and Constraints Extraction} 
When natural language sentences are used to query the trajectory documents, the key topic is to compute the relevance scores (i.e., similarities) between an input sentence and trajectories. The first step is to analyze the sentence and extract important words as query conditions. Given a sentence S, it generally contains the temporal and/or spatial constraint words and their conjunctions. A temporal-constraint word describes a time limitation such as a date or a descriptive word (e.g., morning). A spatial-constraint word describes semantic location information such as a POI name or category. Our query engine extracts the temporal and spatial constraints from the given sentence respectively.

The THU Lexical Analyzer (THULAC)~\cite{li2009punctuation} is employed to split a query sentence into words. THULAC is a word segmentation model and can identify whether a word is a conjunction, a noun or a temporal-related word. The words are automatically divided into conjunctions, temporal-constraint words, and spatial-constraint words.

\noindent \textbf{Temporal Constraints:} 
The temporal-constraint words are directly mapped to a time period or a time stamp. We pre-define several descriptive temporal words and conjunctions to denote temporal constraints, as shown in Table~\ref{tab:Condition}. \zhaosong{The mapping from a word to a time window, such as `morning' to `from 6 am to 9 am', is arbitrarily defined which is easily adjustable by users. This table only enumerates some commonly used words which can be further extended (but may need to change the indexing scheme and query algorithm).}

\begin{table}[tb]
  \caption{Examples of the dictionary of temporal constraints}\label{tab:Condition}
  \scriptsize%
  \begin{tabu}to\linewidth{X[l] X[2.5l]}
  \toprule
  Query Syntax & Description  \\
  \midrule
  morning   & From 6:00 to 9:00 \\
  \midrule
  noon  & From 11:00 to 14:00 \\
  \midrule
  evening   & From 18:00 to 24:00 \\
%  \midrule
%  pass through A   & Trajectory pass a place A \\
  \midrule
  during (T)  & Time conjunction, where T is a time period\\
  \midrule
  A after B & Trajectory go to place 'B' after place `A'. \\
  \midrule
  A before B  & Trajectory go to place 'A' before place `B'\\
  \bottomrule
  \end{tabu}%
\end{table}

\noindent \textbf{Augmented Spatial Constraints:} 
The spatial constraint words in the query sentence can be POI names or descriptive words such as `school'. \zhaosong{Directly using the words to query trajectory documents may not match users' intention. For example, users may use `school' to query middle, high schools, universities, or colleges. Therefore, we apply the word embedding technique to find related words with close meanings.} In particular, the well-known Word2vec~\cite{mikolov2013efficient} algorithm is used to train a word embedding space. The training is applied to all trajectory documents from the given dataset. Moreover, a large corpus Wikipedia data~\cite{rehurek_lrec} is also used to identify widely used common words. 

After training, for an arbitrary query word, its K-nearest neighbors in the trained word embedding space are computed and utilized as augmented spatial constraint keywords for the query. Here, we use the standard cosine similarity to compute the similarity between word $w_1$ and $w_2$:
\begin{equation}\label{equation:simwords}
sim(w_1,w_2)=\frac{\vec{w_1}\cdot \vec{w_2}}{\left | \vec{w_1} \right |\cdot \left | \vec{w_2} \right |},
\end{equation}
where $\vec{w}$ denotes the vector-representation of $w$ in the word embedding space, and $\left | \vec{w} \right |$ denotes its length. 

\subsection{Trajectory Relevance and Retrieval}
\zhaosong{The relevance between the given constraints and the massive trajectory documents are computed in three steps: (1) We first compute a relevance score between each POI and the given spatial constraints. (2) We further integrate the high-level semantics of functional semantics of region to enrich the computation of the relevance scores of POIs. (3) We select the top-K POIs based on the above scores and query the trajectory document database to obtain the query results.}

\subsubsection{Step1: Computing POI Relevance}
Given the augmented spatial constraint keywords, we can find a set of top-K relevant POIs, $L_1,L_2...L_k$. Each POI has a location, description, and category, which also form a POI document. Then this task can be considered as a probabilistic retrieval (PR) problem to find the top-K relevant documents based on a set of keywords. Crestani et al.~\cite{crestani1998document} review a series of studies for solving the PR problem. Robertson et al.~\cite{robertson2009probabilistic} present the Okapi BM25 method to calculate the relevance between the given keywords and documents, \zhaosong{the Term Frequency-Inverse Document Frequency (TF-IDF)~\cite{salton1973specification} measures how important each word is to a document.}

%Topic-based method (e.g., LDA topic model~\cite{wei2006lda}) extracts several topics from the documents and maps each document and the input into a vector. As a result, it is easy to compute the similarity between the input and the document.

%PR problem can be solved by using \zhaosong{the Term frequency-Inverse document frequency (TF-IDF)} method and PL2 method~\cite{amati2002probabilistic, fang2011diagnostic}. 

We use the Okapi BM25 algorithm to compute relevance scores. Given a spatial constraint $W$ which contains $n$ words $w_1,w_2...w_n$, the relevance score $Score(D,W)$ of a POI document $D$ is:
\begin{equation}\label{equation:S1}
Score(D,W)=\sum_{i=1}^{n} R(w_i,D),
\end{equation}
where $R(w_i,D)$ denotes the relevance between $w_i$ and $D$, \zhaosong{which is proportional to the times of $w_i$ appear in $D$, and is offset by the number of POI documents} that contain the word as:
\begin{equation}\label{equation:R}
R(w_i,D) = IDF(w_i)\cdot \frac{TF(w_i)\cdot (k+1)}{TF(w_i) + k\cdot (1-b+b\cdot \frac{\left | D \right |}{avgdl})}.
\end{equation}
Here $\left | D \right |$ denotes the length of the document $D$ in words, and $avgdl$ is the average document length of all the documents. $k$ and $b$ are two parameters which, following the study by Manning et al.~\cite{manning2010introduction}, are set as $k \in \left [ 1.2,2.0\right ]$ and $b = 0.75$. 

Equation~\ref{equation:R} also contains TF-IDF components. Following the study presented in~\cite{robertson2009probabilistic}, we compute $w_i$'s term frequency $TF(w_i)$ by $TF(w_i) = f_{w_i,D}/\left | D \right |$, \zhaosong{where $f_{w_i,D}$ denotes the number of $w_i$ in $D$. $IDF(w_i)$ denotes $w_i$'s inverse document frequency of $D$,} and is computed by:
\begin{equation}\label{equation:IDF}
IDF(w_i) = log\frac{m-M(w_i)+0.5}{M(w_i)+0.5},
\end{equation}
where $m$ is the total number of documents, and $M(w_i)$ is the number of documents that contains word $w_i$.

In this way, the top-K POIs matching the given query sentence in spatial constraints are found.

\subsubsection{Step 2: Enriching POI Relevance with Regional Topics}\label{sec:topic}

\zhaosong{We further enhance the effectiveness of spatial query constraints by taking into account the regional functions when computing the POI relevance. This also provides users more query flexibility in real tasks. For instance, when querying trajectories starting from `restaurants', users may have more interests of restaurants in business districts, or otherwise in transportation areas. Domain users prefer to have such regional attributes in specifying query constraints. Therefore, in computing POI relevance, we need to consider a POI's neighborhood functions, which however cannot be easily addressed due to the complexity of urban POI distribution. Fortunately, there has been data mining work to discover regions of different functions (e.g., ~\cite{Yuan:2012:Discovering}). In our approach, we adopt a similar topic modeling algorithm to compute regional functions (i.e., topics) in each PSR region for the uncertain trajectories. In particular, a set of probabilistic topics are extracted each of which represents a semantic topic of functions (e.g., residential-related, business-related, transport-related, etc.), and each PSR region (Voronoi cell) belongs to these topics with a probabilistic distribution.}

In computation, LDA~\cite{Blei2003Latent} topic modeling is used by regarding Voronoi regions as documents and the inside POIs as words, LDA generates many topics, and each spatial region exhibits several topics which denote their functions. More specifically, given a region $R$, we collect all types of POIs in $R$ and use them as the words in a region document. For instance, a document for the region in the northwest of Figure~\ref{fig:region} contains five POI words:  (restaurant, restaurant, restaurant, restaurant, hotel). After the LDA generates several categories topics, this region can be represented by a probabilistic topic distribution vector $(0.8,0.2)$ in the `Entertainment related' topic and the `Residential related' topic. It shows that this region has the possibility of 80\% and 20\% in the two topics, respectively.

Eventually, the computation of the relevance score of a POI document $D$ to the given words $W$ (Equation \ref{equation:S1}) is enriched as:
\begin{equation}\label{equation:S3}
Score^{'}(D,W) = \alpha \cdot Score(D,W) + \beta \cdot sim_{topic}(\vec{T_L},\vec{T_I}),
\end{equation}
\zhaosong{where $\vec{T_L}$ is the LDA computed probabilistic topic vector of the region containing $D$. $\vec{T_I}$ is the user defined vector to reflect their preferred regional functional topics (see Figure~\ref{fig:conditionspe}). $sim_{topic}(\vec{T_L},\vec{T_I})$ is computed by the standard cosine similarity between $\vec{T_L}$ and $\vec{T_I}$~\cite{erk2012vector,kusner2015word}.}

\zhaosong{As a result, if a POI's document is closely relevant to the spatial constraint words and its surrounding has the user intended region functions, it will obtain a high relevance score by Equation~\ref{equation:S3}. Users can also adjust $\alpha$ and $\beta$ to control the weighted contribution of the keyword matching and the regional topic matching.}

\subsubsection{Step 3: Retrieving Matched Trajectories} \label{sec:retrieval}
The top-K POIs ranked by Equation~\ref{equation:S3} are extracted and used to query the trajectory document database. As discussed in Sec.~\ref{sec:documentindexing}, the temporal-text-indexing structure integrates temporal constraints and inverted keyword indexes of a list of keywords, which can quickly find the matched trajectories. Meanwhile, a relevance score is computed for each retrieved trajectory to the input sentence in the same way discussed above.

Note that, we discuss how to query trajectories based on individual spatial-temporal conditions. When users give a joint condition (e.g., union, intersection, and complement), we perform the query on each condition first and then join them to retrieve the final results.

\vspace{3pt}
\noindent \textbf{Discussion:} The query scheme is implemented by first finding the top-K POIs from the augmented constraints, and then using them to extract trajectories from the trajectory document database. Alternatively, the augmented constraints of keywords can be used directly to query the database. The reasons we use the three-step approach are: \zhaosong{(1) the direct keywords query is slow due to the large data space of massive trajectories. The indexing scheme based on POIs in the database makes it very fast to retrieve results; (2) the direct query results are large and overwhelming, which impedes users' interactive study; (3) the application of Equation \ref{equation:S3} allows users to flexibly control their preference in the query.} Users can study and select top-K POIs (K is controllable), to match and adjust their intention and interactively retrieve trajectories.

\section{Visual Interface}\label{sec:interface}
\zhaosong{We develop a visual interface (Figure~\ref{fig:teaser}) which facilitates users to specify a natural-language-based query and visualize retrieved trajectories within the geo-context, temporal-context, and semantic-context. The interface is shown in Chinese for domain users based on the datasets from a China city, and it is converted to English in this paper and supplemental video.}

\subsection{Query Condition Specification}
\noindent \textbf{Goal:} \zhaosong{This tool should support users to (1) easily add and change natural language query sentences; (2) identify and study the temporal constraints and associated spatial keywords; (3) investigate and manipulate relevant POIs; and (3) control weights of preferred region functional topics.}

\noindent \textbf{Query Input:} As shown in Figure~\ref{fig:teaser} (a), the querying condition specification view includes an input box for users to specify input sentence and to edit input words. For each word, an icon encodes its type and users can change the word type by clicking the corresponding icon and select a type in a drop-down menu (Figure~\ref{fig:teaser} (c)). 

\begin{figure}[tb]
 \centering
 \includegraphics[width=0.95\columnwidth]{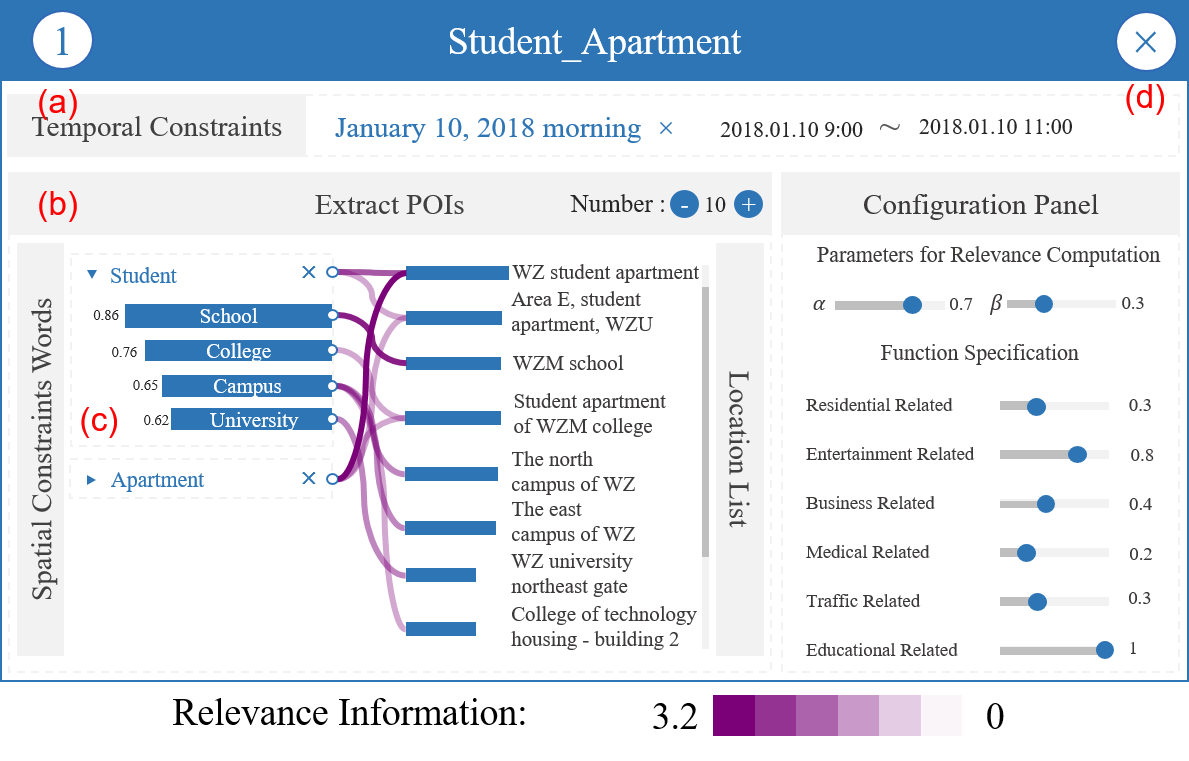}
 \caption{Visual design of the query condition specification. (a) The temporal constraint denotes the time period. (b) The `relevance tree' visualizes the extracted spatial constraint keywords and their relevant POIs. (c) The word `Student' has 4-nearest neighbor words in the word embedding space with their relevance to `Student'. (d) The configuration panel to define weights of region functional topics.}
 \label{fig:conditionspe}
\end{figure}

As shown in Figure~\ref{fig:conditionspe} (a), users can assign temporal conditions by defining the blue text of the temporal-constraint words with the black text of the corresponding period. 

\noindent \textit{Design Rationale:} \zhaosong{In the existing tool in \cite{Aldohuki2016SemanticTraj}, the semantic input is a full freeform sentence since it only handles accurate street and POI names. As we need to process ambiguous location names (e.g., "student"). We enhance the input sentence by allowing users to adjust the word types so that users have more control for their query intention.} Note that our system currently supports queries of a continuous time period. Queries of multiple periods (e.g., every Friday morning) can be achieved by using a time-picker~\cite{chen2018vaud} or Time Wheel~\cite{edsall1997graphical}.

\noindent \textbf{Relevance Tree:} Based on the input, the top-K relevant POI locations are computed, and then users can visually inspect and adjust the extracted POI based query conditions with a `relevance tree' visualization shown in Figure~\ref{fig:conditionspe}(c). Here, two condition nodes on the left represent the two keywords, `student' and `apartment', of the extracted spatial constraints. The related keywords extracted from the word embedding model are shown as bars below each node, whose size shows the histogram of POIs including the corresponding keyword. \zhaosong{Users can hover over each keyword to show the related top $K$ POIs as bars on the right which are sorted by their relevance scores.} The bar size shows the relevance score of a POI to the query condition. The parameter `$K$' (default as 10) can be changed by the plus/minus settings. Moreover, the keyword nodes are linked to the POIs whose name and description contain the keywords by Bezi{\'e}r curves (in Fuchsia). Furthermore, users are allowed to manually assign a word by dragging the word in the input box into a condition node. Users also can delete an extracted word.

\noindent \textit{Design Rationale:} \zhaosong{The design of the relevance tree is inspired by the widely-used, successful Word Tree visualization \cite{Wattenberg2008}. Its node-link based tree display can facilitate the rapid exploration of keywords and related POIs by depicting their relationship to visual cues of font and curve attributes. Its ability to hide/expand a node also helps fight possible cluttering. Alternative designs may utilize word cloud or node-link graph, which however may have an irregular layout comparing with the tree and thus lead to a burden for user interaction.}

\noindent \textbf{Topic Weights:} On the right panel of Figure~\ref{fig:conditionspe}, users can define the preferred topical distribution of surrounding regions according to the region functional topics. We generate 6 topics from the LDA and each topic is assigned with a specific name, color, and icon as shown in Figure~\ref{fig:teaser} (g). Users can specify the weights of these topics on the slide bars. $\alpha$ and $\beta$ can be set by dragging the slide bars as well. In the beginning, $\alpha$ and $\beta$ are set to 1 and 0, respectively. Once these parameters are changed, the relevance scores of POIs are recomputed by Equation~\ref{equation:S3} and the visual interface is refreshed. 

\subsection{Trajectory Visual Exploration}
\noindent \textbf{Goal:} \zhaosong{The queried results should be visually explored with a set of visualization views and interactions. The users should be able to investigate them in the spatial, temporal, and semantic dimensions. The visual interface allows users to perform `overview + drill down' study iteratively.}

%We implemented a result list, a map view, a high-level description projection view, and a timeline view to support the comparison and exploration of the query results.\\

\noindent \textbf{The Detail Result View} displays the queried trajectories (Figure~\ref{fig:teaser} (j)). Each row represents a trajectory labeled with its trajectory ID. The trajectories are sorted by their relevance scores shown in blue bars. Users can check to select them further exploration on the other views.

\noindent \textbf{The Map View} visualizes the result trajectories as polylines in a geographical map (Figure~\ref{fig:teaser} (e)). 
A visual widget (Figure~\ref{fig:teaser} (h)) is designed to interactively specify the visual encodings of the trajectories, such as color and opacity of polylines. Meanwhile, the relevant POIs used to query these trajectories are displayed as markers on the map. 

\begin{figure}[tb]
 \centering
 \includegraphics[width=0.95\columnwidth]{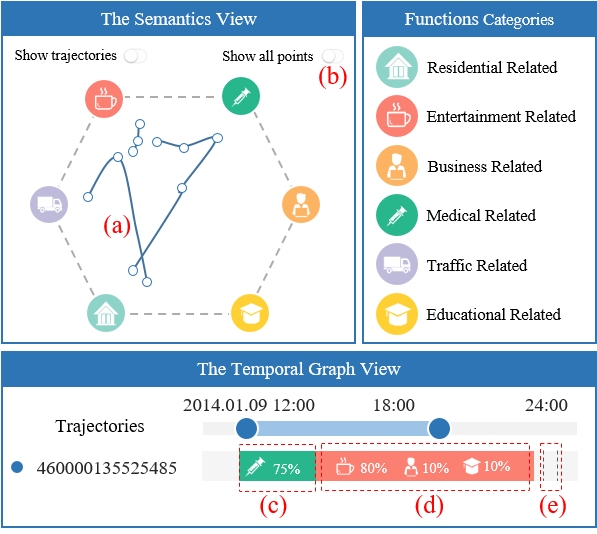}
 \caption{Visual design of the semantics view and the temporal graph view. (a) A polyline shows that one trajectory moves from a `residential-related' region to an `entertainment related' region, then it goes to a `traffic-related' region. (b) The button allows users to show trajectories that meet the constraints to overcome cluttering. (c) The trajectory stays in a region whose dominant functional topic is `medical-related' from `2014.01.09 12:00' to `2014.01.09 14:00'. (d) The trajectory stays in a region with different possibilities in multiple topics: `entertainment-related 80\%, business-related 10\%, educational-related 10\%' from `2014.01.09 14:00' to `2014.01.09 21:00'. (e) The trajectory passes a region in a short time without showing its functional information.}
 \label{fig:projection}
\end{figure}

\noindent \textbf{The Semantics View} visualizes the trajectories as polylines inside a topic hexagon (Figure~\ref{fig:projection}).
\zhaosong{Each vertex represents a regional functional topic which is depicted by an intuitive icon. Users can interactively drag these vertices to change their radial locations.} Inside this hexagon, each selected trajectory is visualized while the points of this trajectory are drawn close to their related topic vertices. Here, a force-directed layout algorithm is employed to adjust these point positions iteratively. Specifically, we first find the spatial region that contains one point and then get its LDA-based topical vector $\vec{V}$. The value of $\vec{V}$ defines the attractive force between the topic vertices and this trajectory point. As a result, the closer a trajectory point is located to a vertex, the more possible its surrounding region has the corresponding topic. In this way, users can easily find the functional information of the regions that a trajectory has passed. 

Figure~\ref{fig:projection} shows an example with three trajectories. Most of their points are located in regions whose dominated functional topic is `entertainment-related'. Users can inspect details of individual trajectories. For example, the trajectory is shown in Figure~\ref{fig:projection} (a) moves from a `residential-related' region to an `entertainment related' region, then it goes to a `traffic-related' region. To overcome possible cluttering, users are able to only display the trajectory points that meet the query constraints (Figure~\ref{fig:projection} (b)).

\noindent \textit{Design Rationale:} \zhaosong{This semantic view abstracts and visualizes individual trajectories in a `region functional topic space' which is intuitive and provides easy interaction. It helps users immediately identify their main mobility patterns among city regions. Instead, if such trajectory topical information is visualized over the map view, it will be hard for users to find such a pattern due to the overlapped geo-information and also hard to compare different trajectories.} 

\noindent \textbf{The Temporal Graph View} visualizes trajectories as colored bars over a time axis (Figure~\ref{fig:teaser} (i)). Here each trajectory bar represents a trajectory consisting of several colored segments. Each trajectory segment is colored to denote the dominant functional topic of its located region while an icon is added to further enhance understanding. Moreover, a time slider is used on the time axis to adjust the width and position of each segment. For example, the line in Figure~\ref{fig:projection} shows that a person stays in a `medical-related' region between ``12:00'' to ``14:00'', and then goes to an `entertainment-related' region. There are three types of visualized segments. The first type (Figure~\ref{fig:projection} (c)) only shows the dominant functional topic of its located region.
The second type sequentially lists other possible topics. For example, other functional topics of Figure~\ref{fig:projection} (d) are `business related' and `educational related'. The third type (Figure~\ref{fig:projection} (e)) uses a grey segment hide the functional topics. The three types are automatically selected by checking whether the segment width is enough to show the necessary icons.

\noindent \textit{Design Rationale:} \zhaosong{We adopt this design because such visualization has proven to be useful in analyzing chronological sequence data~\cite{zhang2014visual, guo2017eventthread}. It allows users to easily adjust and zoom in/out to focused time periods. The extra information about topics is easily shown on the segments.}

%begin{figure}[tb]
% \centering
% \includegraphics[width=0.95\columnwidth]{pictures/semantimeline.png}
% \caption{Visual design for a trajectory: }
% \label{fig:semantimeline}
%\end{figure}

\noindent \textbf{Coordinated Interactions:} All views are coordinated by interactions. In the query condition specification view, when a POI is hovered, its location and details are also displayed in the map view. Furthermore, when a keyword is hovered, all related POI locations and their details are shown on the map. When $\alpha$ and $\beta$ or high-level description information is modified, trajectories in the detail result view are refreshed accordingly.

In the map view, users can zoom in and out to study details. Meanwhile, users can highlight any trajectory by hovering it in the detail result view. In the detail result view, users can check the selection box to add arbitrary trajectories to the temporal graph view and the semantics view. In the temporal graph view, when a trajectory segment is hovered, both the trajectory and the trajectory point are highlighted in the map view and the semantics view. Moreover, a time period can be specified by setting the time slider so that all trajectories within the period are filtered and shown on the map. Furthermore, users can directly draw regions on the map so that the trajectories passing it are filtered and displayed.

\section{Case Studies}\label{sec:casestudy}
\noindent \textbf{Trajectory Dataset:} We demonstrate our system by applying the query engine and visual interface on a real-world human trajectory dataset in a city with about 14 million citizens. It is collected between Jan. 10th, 2014 and Jan. 31st, 2014 and its total size is 736 GB. The datasets include:
\begin{itemize}
    \item POIs: There are 862,635 POIs, each of which records its GPS coordinates, name and type.
    \item Mobile phone trajectories: There are 3 billion records of 7 million mobile phone users. Each record of a sampling point contains the anonymized User ID, the cell phone base station ID, and the timestamp. The location accuracy varies from 500 to 5000 meters depending on the coverage of base stations.
\end{itemize}

\noindent \textbf{Visual Query Tasks:} 
\zhaosong{In designing our prototype, we establish several specific tasks by collaborating with a domain expert from the Urban Planning College of a local university of the city where the dataset is collected. In this paper, we show two usage cases of trajectory analysis proposed by the expert in his real tasks.
The first case focuses on exploring tourist mobility patterns related to the city traffic since the tourists greatly contribute to the city's transportation issues. The second case intends to analyze the daily mobility features of local students, as the expert claims that ``providing students with convenient transportation services has always been one of the important tasks of urban traffic management''.}

\zhaosong{The cases are presented to illustrate the functions of our approach: the first case shows how our approach supports intuitive natural language query, interactive condition specification, and flexible trajectory exploration; and the second case demonstrates how our system can help analyzers generate questions and form insights during the exploratory process.}

\zhaosong{The two cases only provide a starting point for identifying and performing visual query tasks involving uncertain trajectories. In general, human trajectories have a large variety of applications in the areas of urban planning, transportation, environment, criminology, and so on \cite{zheng2014urban}. Visual analytics tasks of trajectories have been well studied and surveyed \cite{chen2015survey,andrienko2017visual}. Please note that this paper focuses on data pre-processing and visual querying but not on the post-analytics and data mining. However, most of these tasks can be supported by our query engine for uncertain human trajectories. We will conduct comprehensive application usages of this prototype with multidisciplinary experts, based on which we will try to formalize their requirements to form a roadmap of assessing uncertain human trajectories. These will be the major future directions of our work.} 

\noindent \textbf{Urban Regions and Functional Topics:} 
As discussed in our algorithms, we partition the city into Voronoi regions of base stations. Then, we extract a set of region functional topics from POI data by the LDA model based on the whole city's POI data. \zhaosong{The domain expert suggested that we set the number of topics between 5 to 10. The reason is that if there are too many topics, each topic will only represent one or two types of POI and they are too detailed to investigate. On the other hand, if the number of topics is too small, most regions will have the same dominating topics and it is difficult to distinguish them by their functions. We generated the results where the number of topics is from 5 to 10. After a few experiments, the expert empirically selected 6 topics in our prototype, and the keywords list in these topics are adjusted with minor manual work.} Each topic is assigned with a specific name and icon as shown in Figure~\ref{fig:teaser} (g).

%In practice, we empirically set the semantic topic number to be 6 after several tests. Each topic is assigned with a specific name and icon corresponding to a type of functionality area as shown in 

\subsection{Examining Tourist Traffic Pattern}
The goal is to analyze the trajectories of tourists especially those who went to the popular tourist attractions in the downtown area. 
As shown in Figure~\ref{fig:teaser} (a), the expert first defines a query sentence as "Query trajectories passed through tourist attractions during January 25, 2014". Figure~\ref{fig:teaser} (b) displays the extracted spatial constraint of keyword `tourist attractions'. It lists twenty POIs related to `tourist attractions'. Because popular tourist areas in downtown usually contain entertainment-, business-, and traffic-related POIs, the expert modify the weights of these region functional topics as $1$, $0.6$ and $0.6$, respectively (others as $0$). Then, the expert adjusts $\alpha$ and $\beta$ to give different weights of the keyword relevance and the preferred region functions. In this example, by setting $\alpha = 0.6$, $\beta = 0.4$ to perform the query, \zhaosong{the expert gives a little more importance to the keyword relevance. The system returns a significant amount of trajectories for next step exploration.}
\begin{figure*}[tb]
 \centering
 \includegraphics[width=\linewidth]{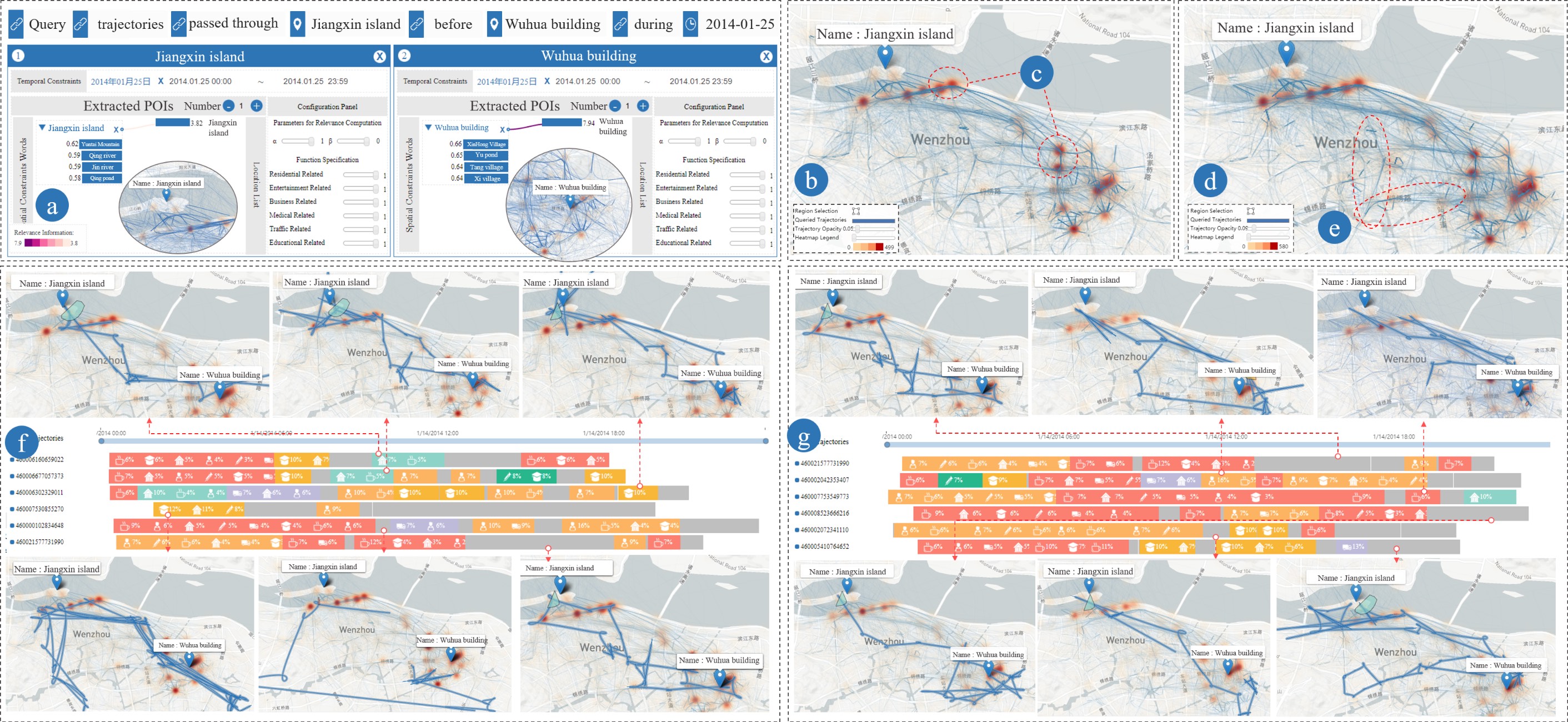}
 \caption{Examining tourist traffic patterns. (a) The extracted spatial-temporal constraints. The queried trajectories should pass them one by one. (b) The queried trajectories from `Jiangxin island' to `Wuhua building'. (c) The heatmap displays two main commuting routes sharing the same crossroads. (d) The queried trajectories from the `Wuhua building' to `Jiangxin island'. (e) The commuting route in the middle of the city. (f) The temporal graph view visualizes top-6 trajectories of (b), where the first and second trajectories stay in `Jiangxin island' for several hours, the fourth trajectory travel along the same route many times, and others pass through the regions without staying. (g) The top-6 relevant trajectories of (d), where the second and the third trajectories stay in `Wuhua Building' during day time and others pass `Jiangxin island' without staying.}
 \label{fig:case12}
 \vspace{-5mm}
\end{figure*}

\noindent \textbf{Overview of Trajectories:} The expert first browses the map view and tries to figure out the most popular tourist attraction in the city. The heatmap shows that the queried trajectories are mainly distributed in the northwest, the middle, and the east of the city (Figure~\ref{fig:teaser} (e)), where the expert also displays the relevant POIs.

Then, he filters the trajectories passing tourist attractions from `9:00 am' to `6:00 pm' by modulating the time slider. There remain approximately six thousand trajectories (Figure~\ref{fig:teaser} (j)).
The expert further inspects the POIs in the northwest and eastern regions and finds two popular tourist attractions named `Jiangxin island' and `Wuhua Building'.
The expert draws a region to filter only those trajectories passing the island (Figure~\ref{fig:teaser} (f)). He finds that most trajectories landed from the east side of the island. There are no bridges and a ferry helps tourists get on the island, whose capacity may need to be increased to improve urban traffic.

\noindent \textbf{Spatial-temporal and Semantic Exploration:} For a drill-down study, the expert form a query sentence as `Query trajectories passed through Jiangxin island before Wuhua Building during January 25, 2014' to analyze the commuting between the island and a city landmark Wuhua Building.
By screening the condition specification view  (Figure~\ref{fig:case12} (a)), he realizes that there are two spatial constraint words that are extracted, and their order denotes that the trajectories should pass through them one by one. He sets the number of extracted POI in each node as `1' and performs the query. As shown in Figure~\ref{fig:case12} (b), 415 trajectories are queried. In the map view, the experts modify the opacity of trajectories and observe two main commuting paths both along the river across the city. These trajectories share the same crossroads (Figure~\ref{fig:case12} (c)) which are important traffic hubs in the city. \zhaosong{Experts suggest that if we want to maintain a smooth transit between the two tourism locations, we should make a high priority to manage the traffic of these hubs.}

The expert further selects the top-6 relevant trajectories and studies them in the temporal graph view (Figure~\ref{fig:case12} (f)). He observes that the first and second trajectories travel to `Jiangxin island' from `Wuhua Building' in the morning, then they stay there for several hours and then return to `Wuhua Building' at night. The fourth trajectory moves along the same bus route many times which may be from a driver. The other trajectories pass through these two regions without staying. Such detail studies can be of importance to identify typical or suspicious movement behaviors.

\noindent \textbf{Condition Specification:} 
The expert then changes the order of the spatial constraint words by dragging them directly. 474 trajectories are queried, which travel to `Wuhua Building' before `Jiangxin island'. From the map view (Figure~\ref{fig:case12} (d)), he finds that there are three popular commuting roads. Besides the two riverside roads, the third route lies in the middle of the city downtown (Figure~\ref{fig:case12} (e)).
The expert again selects top-6 relevant trajectories. From the temporal graph view and the map view (Figure~\ref{fig:case12} (g)), he realizes that the second and the third trajectories went to the `Wuhua Building' at the day time and went to `Jiangxin island' at night. Other trajectories pass through `Jiangxin island' without staying. 

\subsection{Exploring Student Related Trajectories}
\begin{figure*}[tb]
 \centering
 \includegraphics[width=\linewidth]{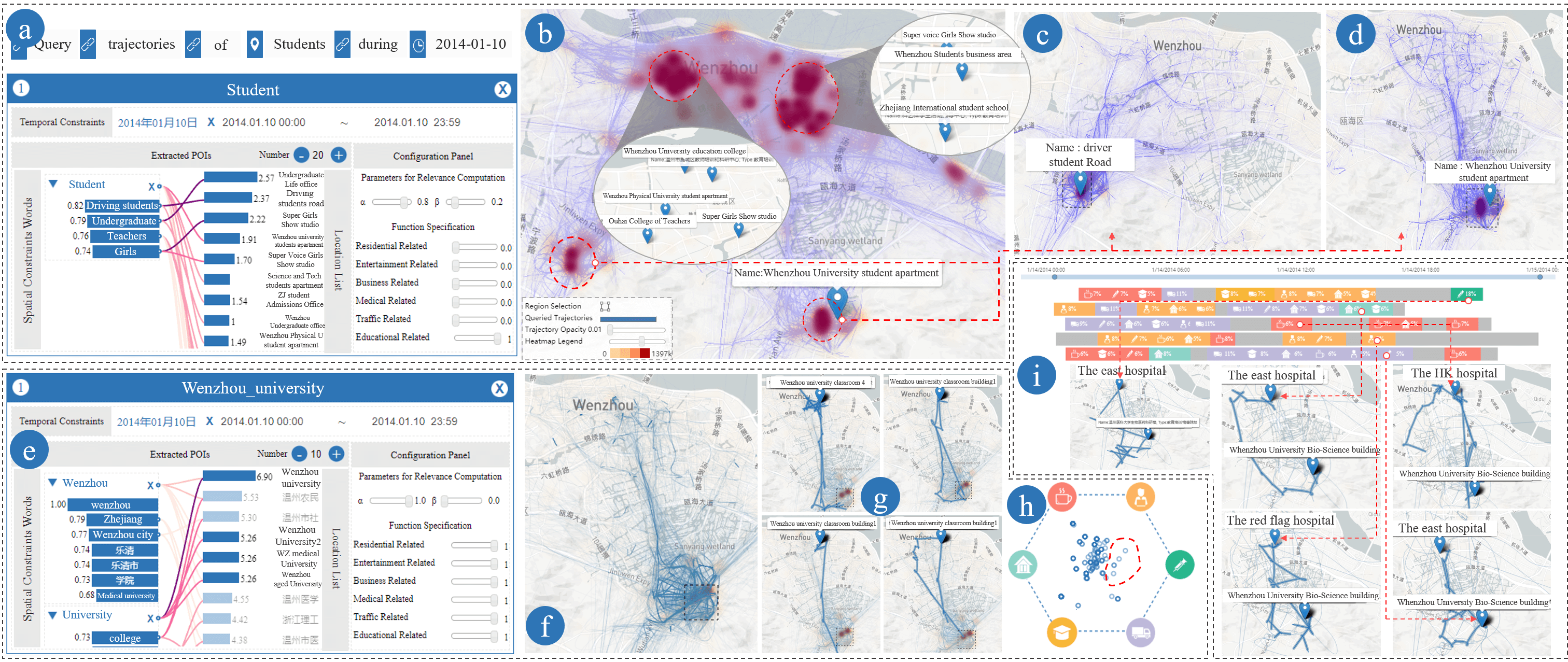}
 \caption{Exploring student related trajectories. (a) The input box shows a query sentence `Query trajectories of students during 2014-01-10'. (b) The map view shows four hot regions colored in a dark red. (c) The trajectories pass through the `student road'. (d) The trajectories pass through the `Wenzhou university student apartment'. (e) Spatial keywords and relevant POIs including four campuses of `Wenzhou university'. (f) The trajectories pass through the south campus. (g) The trajectories pass the downtown campus during day time and return to suburb south campus at night. (h) The semantics view shows that there are many trajectory points close to the `medical-related' topic. (i) The trajectories depart from the campus and visit a hospital in the downtown area.}
 \label{fig:case2}
 \vspace{-4mm}
\end{figure*}

\zhaosong{The expert wants to study students' trajectories which is a broad goal. In the beginning, he does not have fixed location names to perform the query.} He thus enters an input sentence as `Query trajectories of students during Jan. 10 2014' (Figure~\ref{fig:case2} (a)). Note that such a query of human trajectories is only possible by our new natural language query engine. Moreover, the expert modifies the weight of the `educational-related' topic as 1, and others as 0. He sets $\alpha = 0.8$, $\beta = 0.2$ to perform the query. \zhaosong{Here $\beta = 0.2$ so that regional topics are less important than keyword matching. Therefore, student trajectories that are not located in an educational-focused region tend to be extracted which is of the expert's interest.}

\noindent \textbf{Overview of Trajectories:} The queried trajectories are shown in Figure~\ref{fig:case2} (b). The expert observes that there are four hot regions colored in a dark red. He focuses on the southwest and south of the city since they are far away from downtown. He checks their names and filters the trajectories passing them. The southwest region shows the`student road' of a driver school (Figure~\ref{fig:case2} (c)), whose trajectories are mainly distributed on a highway. The south region (Figure~\ref{fig:case2} (d)) is around `Wenzhou university student apartment' and some of its trajectories are located in the downtown area. Subsequently, the expert conducts the analysis at this university.

\noindent \textbf{Spatial-temporal and Semantic Exploration:} He changes the query input to `Query trajectories that passed through Wenzhou University during Jan. 10 2014' and performs the query. By hovering over the word `university', he finds that there are several campuses of Wenzhou University (Figure~\ref{fig:case2} (e)). Most of the campuses are located in the downtown area. Then, he defines a region on the map to analyze the trajectories of the south campus.
As shown in figure~\ref{fig:case2} (f), most of the trajectories are located around this campus. However, there are several trajectories visiting downtown. The expert displays these trajectories on the map (Figure~\ref{fig:case2} (g)). He finds that they visit the downtown campus during day time and return to the south campus at night.
Then, he selects these trajectories in the detail result view and adds them to the semantics view (Figure~\ref{fig:case2} (h)). In this view, many trajectory points are close to the `medical-related' topic. He then changes the query input to `Query trajectories that passed through Wenzhou University and hospital during Jan. 10 2014'. As shown in Figure~\ref{fig:case2} (i), these trajectories depart from the university to a hospital. This finding indicates that students of the south campus prefer to visit the hospital in the downtown area. More medical services may be provided to the south campus area. 

%From the semantic view, we find that the major topic is ``Hospital" which indicate the students prefer to visit hospital in the downtown area instead of in the campus (\ref{fig:case3} (e)).

%The analyst also studies trajectories whose semantics include  ``education". The queried trajectories mainly locate in the campus. This implies that there are many facilities on education around the campus (\ref{fig:case3} (g)).

%Finally, the analyst identifies a trajectory cluster that contains the topic ``downtown". By further specifying  ``pass college and pass train station during [Jan. 10 2014]" and performing querying, it is found that these trajectories finally end in the campus and leave the city(\ref{fig:case3} (f)). 

\section{Discussion} \label{sec:discussion}

\subsection{Spatial and Temporal Uncertainty}
\zhaosong{Uncertainty in trajectories conceptually can have two types: spatial uncertainty and temporal uncertainty. Arising from sampling inaccuracy and privacy protection, spatial uncertainty is a major topic in mobility data processing and analysis \cite{fisher1999models,duckham2006location,andrienko2017visual}. Our approach presents a visual query solution with the POI-based textualization and natural language search engine. Temporal uncertainty of trajectories, although not happen as often as spatial uncertainty, may also arise from the inaccurate or lack of time information at sampling points. It is more complex to handle since the temporally continuous trajectories may not be formed correctly so that the data management and visual analysis are not directly plausible. Our approach does not handle this complexity which is a future topic. In this paper, we allow users to define temporal constraints with descriptive words that link to adjustable time periods, while we assume the timestamps are correct. In future work, we may develop a temporal uncertainty identification algorithm to give the questionable points a temporal possible range. Then, the relevance computation will need to incorporate this range in finding close POIs with a probability. We will also improve the definition and understanding of temporal query words in natural sentences.}

\zhaosong{Moreover, our approach handles spatial uncertainty mainly by linking and textualizing an uncertain trajectory point to the POIs in its PSR (possible spatial region). The region functional topics are also discovered for these PSRs. With the mobile phone based human trajectory data in this paper, PSRs are implemented as the Voronoi cells subdividing urban space based on given base stations. This approach can be extended to other types of uncertainty trajectories (e.g., passenger trajectories who used metro, bus~\cite{zeng2013visualizing} or taxi~\cite{Huang2015TrajGraph}) by defining different kinds of PSRs. For example, inaccurate sampling points on taxi trajectories may link to surrounding circular (or polygonal) PSRs with probable sizes or ranges. This strategy is used in mining functional regions from taxi trajectories \cite{Yuan:2012:Discovering}. Alternatively, PSRs can also be implemented by partitioning a city into small regions by other means (e.g., census regions~\cite{huang2019exploring}). Consequently, our documentation and search engine can be applied in a similar way to mobile human trajectories.}

\subsection{Performance and Scalability}
Query performance is one of the key focuses when designing the query engine and visual interface over massive trajectories. \zhaosong{First, as shown in Sec. \ref{sec:documentindexing}, we design the specific indexing scheme, and consider efficient data partitions together with disk I/O (similar to Nanocubes~\cite{lins2013nanocubes}), to support real-time data access from the database. Second, we apply the three-step query process as discussed in Sec. \ref{sec:retrieval}. Third, to visualize the trajectories and their semantics, we follow the mantra `Overview first, zoom and filter, then details-on-demand'~\cite{shneiderman2003eyes} in visual interface design. As a result, massive human trajectory data with 3 billion records and 7 million mobile users in one month (736 GB) has been smoothly queried and visually explored in our prototype.} Table~\ref{tab:performance} summarizes the query times in our test of the query efficiency, where three typical queries are applied:
\begin{itemize}
    \item Task 1: Query trajectories pass through `WZ train station' in 2 hours (`2014.01.10 10:00' to `2014.01.10 12:00').
    \item Task 2: Query trajectories pass through `WZ train station' in 2 days (`2014.01.10 00:00' to `2014.01.11 24:00').
    \item Task 3: Query trajectories move from `Baihua park to `WZ train station' in 2 hours (`2014.01.10 10:00' to `2014.01.10 12:00').
\end{itemize}

\begin{table}[t]
  \caption{Query Performance for Three Tasks}\label{tab:performance}
  \scriptsize%
    \centering%
  \begin{tabu}to\linewidth{X[2l] X[l] X[l] X[l]
    }
  \toprule
   Task & Task1 (sec) & Task2 (sec) & Task3 (sec)  \\
  \midrule
  Data Access Time & 2.07 &  13.48 &  4.35  \\
   \bottomrule
  \end{tabu}%
\end{table}

\zhaosong{From the result, we find that the query performance is proportional to the queried time period since more data partitions and trajectories exist for the query engine to fetch. 
The selection of time parameters affects the query efficiency. A small time window size leads to more data partition blocks and using a large time window size stores too much data within one data block. In future work, we may analyze the query tasks from real users and then optimize the data block size.}

\zhaosong{When the datasets become even bigger, such as multiple months or years of human trajectories, the system needs to be extended to a distributed computing platform. We will extend the database and text search engine to a parallel database and computing schemes. Fortunately, many database tools such as Postgresql, MongoDB and Solr now support full-text search and indexing in distributed servers, which will be adopted in our extension.}

\section{User Feedback} \label{sec:feedback}
\zhaosong{The prototype was assessed by the domain expert from the urban planning college and about 10 graduate students in CS of a local university. After freely using the system, they provided us preliminary feedback about the natural language based visual interface.} 

\zhaosong{Overall, the users were satisfied with our natural language based input method. One user claimed `this is indeed a user-friendly query method, I can easily find the trajectories I want.' In particular, users were interested and impressed by the keywords extraction and automatic association to relevant words. The expert said `I was very surprised to find that no matter what word I typed, the system was always able to find several relevant locations.' Other users said `It seems that the system can understand the meaning of the words which is different from a keyword query. In the system, I can clearly view how my inputting words are transformed into locations from the above two views.'}

\zhaosong{They also pointed out some limitations. While our system is designed for exploration of spatial uncertain trajectories, it was suggested that the spatial attributes of trajectories can be presented more clearly, since sometimes the map view of many query results became hard to understand and required lots of interactions. A possible solution is to aggregate the trajectories by edge bundling algorithm~\cite{ersoy2011skeleton}. Another limitation is `Now, It is time-consuming to analyze the trajectories one by one. If the system can automatically find the similar trajectories and group them, it will be convenient to analyze a large number of trajectories.' This can be solved by providing automatic clustering and recommendation to the query results. As our current work focusing on queries, we will conduct such analytics work over query results in our future work.}

\zhaosong{A comprehensive and formal user study is demanded to evaluate and improve the system. Next step, we plan to work with different domain experts in multiple applications for this purpose.}
\section{Conclusion and Future Work}\label{sec:conclusions}
We present a query engine to convert, store and retrieve spatial uncertain trajectories via intuitive natural-language input. A visual analytic system helps analysts explore the trajectories of millions of mobile phone users. This approach allows users to easily specify data query conditions, adjust them interactively, and finally explore the query results with a set of visual tools and effective interactions. We demonstrate the functionality and usage of our approach through two cases which based on a real-world human trajectory dataset.

We have discussed some limitations and future work in Sec. \ref{sec:discussion} and Sec. \ref{sec:feedback}. In addition, the ability to understand all kinds of natural language queries is limited. For instance, Boolean query conditions are not conducted currently. The challenge lies in finding the conjunction and set the relationship between the conditions. A possible solution is to pre-define several conjunctions and query models over them. Second, our system supports a detail study of the semantics of a small number of trajectories. \zhaosong{The semantic visualization method for massive trajectories (e.g., hundreds or thousands of trajectories) is still lacking. This is a challenging task considering the limited visual estate and human perception capability. We may design an interactive drill-down query model based on natural languages to solve the problem with new interactions.}

\acknowledgments{This work was supported in part by National Natural Science Foundation of China (U1609217, 61772456, 61761136020). Y. Zhao's work was supported in part by the U.S. NSF grants 1535031, 1535081, and 1739491.}

\bibliographystyle{abbrv-doi}
\bibliography{template}

\end{document}